\documentclass[twocolumn]{aastex6}
\usepackage{epsfig}
\usepackage{graphicx}
\usepackage{float}
\usepackage{amsmath}
\usepackage{color}
\usepackage{amssymb}
\usepackage{amsfonts}
\usepackage{units}
\usepackage{bm}

\def\be{\begin{eqnarray}}
\def\ee{\end{eqnarray}}

\begin{document} 

\title{Pair separation in parallel electric field in magnetar magnetosphere and narrow spectra of Fast Radio Bursts}

\author{Yuan-Pei Yang\altaffilmark{1}, Jin-Ping Zhu\altaffilmark{2,3}, Bing Zhang\altaffilmark{4} and Xue-Feng Wu\altaffilmark{5}}

\affil{
$^1$ South-Western Institute for Astronomy Research, Yunnan University, Kunming 650500, Yunnan, P.R.China; ypyang@ynu.edu.cn;\\
$^2$Kavli Institute for Astronomy and Astrophysics, Peking University, Beijing 100871, China;\\
$^3$ Department of Astronomy, School of Physics, Peking University, Beijing 100871, China \\
$^4$ Department of Physics and Astronomy, University of Nevada, Las Vegas, NV 89154, USA; zhang@physics.unlv.edu\\
$^5$ Purple Mountain Observatory, Chinese Academy of Sciences, Nanjing 210023, China
}

\begin{abstract} 
When the magnetosphere of a magnetar is perturbed by crustal deformation, an electric field $E_\parallel$ parallel to the magnetic field line would appear via Alvf\'en waves in the charge starvation region. The electron-positron pair bunches will be generated via two-stream instability in the magnetosphere, and these pairs will undergo charge separation in the $E_\parallel$ and in the meantime emit coherent curvature radiation. Following the approach of \cite{yan18}, we find that the superposed curvature radiation becomes narrower due to charge separation, with the width of spectrum depending on the separation between the electron and positron clumps. This mechanism can interpret the narrow spectra of FRBs, in particular, the spectrum of Galactic FRB 200428 recently detected in association with a hard X-ray burst from the Galactic magnetar SGR J1935+2154.
\end{abstract} 


\section{Introduction}

Fast radio bursts (FRBs) are mysterious radio transients with millisecond durations and extremely high brightness temperatures from cosmological distances \citep{lor07,tho13,cha17,ban19,pro19,rav19,mar20}. Recently, an FRB-like event (FRB 200428) with two peaks separated by $30~{\rm ms}$  \citep{CHIME2020,STARE2020} was detected from the Galactic magnetar, SGR J1935+2154, during its active phase in association with a hard X-ray burst \citep{HXMT2020,INTEGRAL2020,Konus2020,AGILE2020}.
The Canadian Hydrogen Intensity Mapping Experiment (CHIME) detected FRB 200428 at $(400-800)~{\rm MHz}$ with a dispersion measure ${\rm DM}=333~{\rm pc~cm^{-3}}$ and a fluence reaching a few hundreds kJy ms \citep{CHIME2020}. Meanwhile, the Survey
for Transient Astronomical Radio Emission 2 (STARE2) reported the simultaneous detection of one of the two peaks (likely the second peak) of FRB 200428 with an extremely large fluence reaching $\sim 1.5~{\rm MJy~ms}$ at $1.4~{\rm GHz}$, which is about 40 times less energetic compared with the weakest extragalactic FRBs observed so far \citep{STARE2020}.
The associated hard X-ray burst was detected by Insight-HXMT \citep{HXMT2020}, INTEGRAL \citep{INTEGRAL2020}, Konus-Wind \citep{Konus2020} and AGILE \citep{AGILE2020}. In particular, there are two hard X-ray peaks whose arrival times are consistent with the two FRB peaks after de-dispersion \citep{HXMT2020,Konus2020}.

Although FRB 200428 was found to be associated with a hard X-ray burst, deep searches by Five-hundred-meter Aperture Spherical Telescope (FAST) for FRBs revealed no single detection, even during the epochs when 29 soft-$\gamma$-ray bursts were detected by Fermi GBM \citep{FAST2020}. This suggests that the FRB-SGR association is very rare.
Among other possibilities, the low probability of association could be due to the narrow spectra of FRBs \citep{FAST2020}. Such narrow spectra have been hinted by the extreme variation of spectral indices among different bursts of FRB 121102 \citep{spi16} as well as the relative fluence of the two peaks  of FRB 200428 as observed by CHIME and STARE2.

The association between FRB 200428 and the two hard spikes of the X-ray burst from SGR J1935+2154 suggests that they very likely share the same origin. The high-energy emission from a magnetar is widely interpreted as due to a magnetospheric activity \citep{tho05,bel07}. 
When a magnetar magnetosphere is trigged by crustal deformations, an electric field $E_\parallel$ parallel to the magnetic field line would appear via Alvf\'en waves in the charge starvation region \citep{kum20,lu20}. The electron-positron pair bunches will be generated via two-stream instability in the magnetosphere, and these pairs will undergo charge separation in the $E_\parallel$ and emit coherent curvature radiation.
In this work, we calculate the coherent curvature radiation spectrum of spatially separated pairs, and apply it to the observed spectra of FRB 200428.
The paper is organized as follows.
We first discuss the FRB generation mechanism within the magnetosphere of a magnetar in Section \ref{magnetosphere}. We then calculate the coherent curvature radiation spectra of the separated pair clumps in Section \ref{spectrum}. 
The results are summarized in Section \ref{discussion}. The convention $Q_x\equiv Q/10^x$ is adopted in cgs units.

\section{FRBs from Magnetosphere Activities}\label{magnetosphere}

Various FRB models can be divided into ``far-way'' models and ``close-in'' models based on the distance of the emission region from the neutron star \citep{lu20}. The former suggests that the energy is dissipated via an outflow interacting with the ambient medium, and radio emission is produced by certain synchrotron maser mechanisms \citep{lyu14,wax17,bel17,bel19,met19,mar20b,yu20}. 
The latter suggests that the radio emission is from the magnetosphere of a neutron star \citep{pen15,cor16,kat16,kum17,zha17b,lu18,yan18,kum20,wwy20,lu20,dai20,wjs20}.
We believe that a magnetospheric origin of FRB emission is most likely, based on the following observational evidence or theoretical arguments. The issues of synchrotron maser model to interpret FRB 200428 has been discussed by \cite{lu20} (cf. \citealt{mar20b}).
\begin{itemize}
	\item The two pulses of FRB 200428 \citep{CHIME2020} were associated with two hard spikes of the hard X-ray burst from SGR J1935+2154 \citep{HXMT2020,Konus2020}. The high-energy emission of SGRs has been widely believed to be caused by the magnetosphere activity of the magnetars \citep{tho05,bel07}. It is most natural to attribute the radio emission also from the magnetosphere \citep{HXMT2020}.
	\item Several magnetars have been identified as pulsed radio emitters \citep{cam06,cam07}, e.g., XTE J1810-197 and 1E 1547.0-5408. The coherent radio emission of these magnetars is well consistent with due to a magnetospheric origin \citep{wan19c}. 
	\item The observation of the frequency drift of FRB 121102 is $\dot\nu\sim(10-10^3)~{\rm MHz~ms^{-1}}$ at $\nu\sim1~{\rm GHz}$ \citep{hes19}. This information may be used to estimate the size of the emission region
	\be
	r\sim\frac{c\nu}{\dot\nu}\lesssim10^9~{\rm cm},
	\ee
	which is smaller than the light cylinder $R_{\rm LC}=cP/2\pi\simeq5\times10^9~{\rm cm}~P_{0}$ of a neutron star.  Indeed, such a drifiting behavior can be well interpreted within the framework of magnetospheric coherent curvature radiation models \citep{wan19}.
\end{itemize}

\begin{figure}
    \centering
	\includegraphics[width = 1\linewidth , trim = 70 80 100 60, clip]{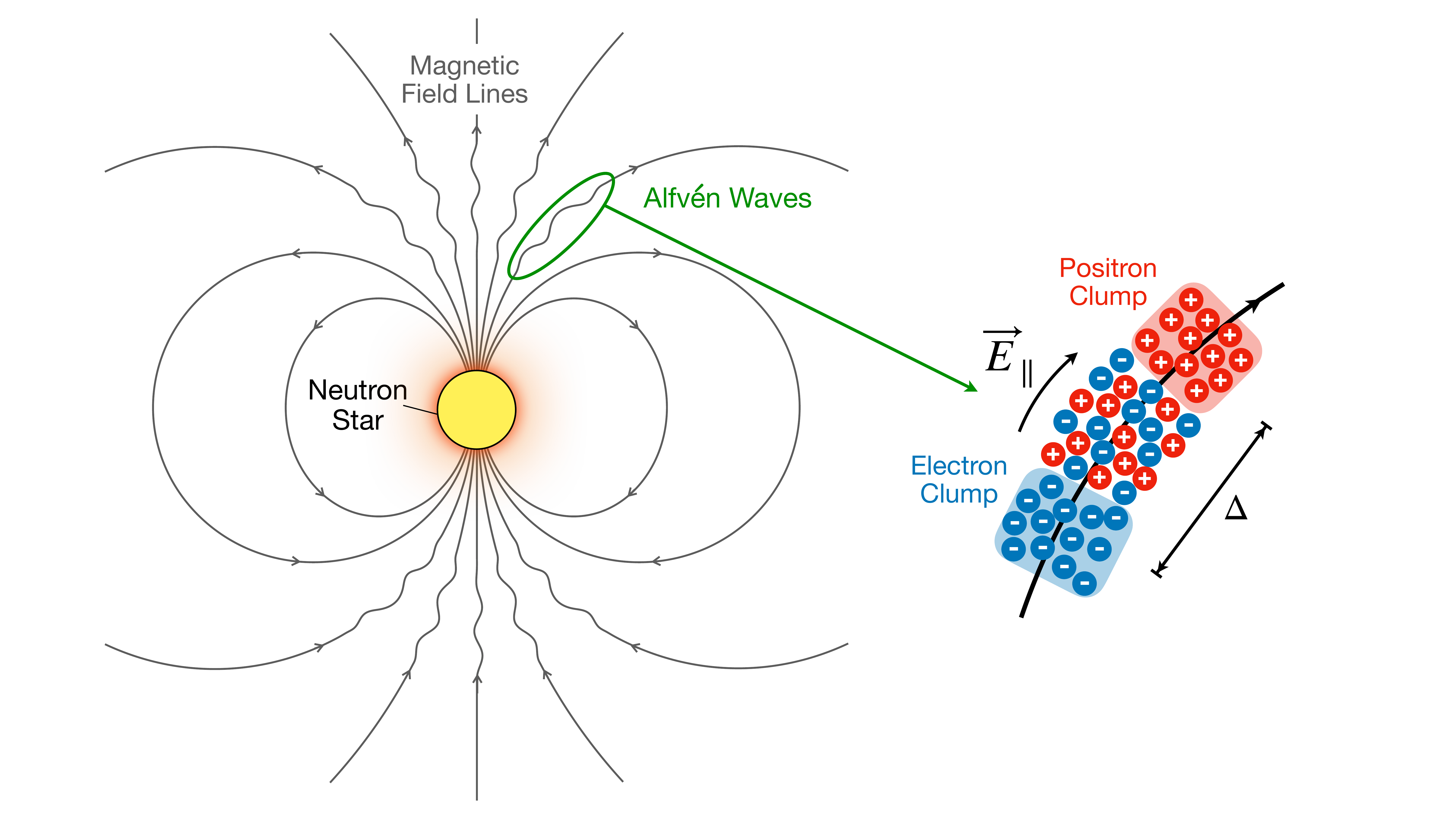}
    \caption{The cartoon picture of the magnetospheric FRB model (see also \cite{lu20}). The green ellipse denotes the region where $E_\parallel$ is developed and charges separate. The red points denote leading positron clump, and the blue points denote the trailing electron clump. $\Delta$ is the separation between the clump pair. The $E_\parallel$ is triggered by Alvf\'en waves reaching the charge starvation region \citep{kum20}.}\label{fig1}
\end{figure}

Coherent curvature radiation by bunches could be an attractive mechanism to generate FRBs from the magnetosphere of a magnetar \citep[e.g.,][]{kum17,yan18,lu20}. For an FRB at $\nu\sim1~{\rm GHz}$, the electron (positron) Lorentz factor is required to be
\be
\gamma=\left(\frac{4\pi\rho\nu}{3c}\right)^{1/3}\sim520\rho_9^{1/3}\label{gamma}
\ee
where the curvature radius is about $\rho\sim 4r/3\theta$, and $\theta$ is the poloidal angle. The rapid cooling of the leptons in the coherent bunch demands that there should be an electric field ($E_\parallel$) along the magnetic field lines to continuously provide emission power \citep{kum17}. In non-twisted pulsar magnetospheres, such an $E_\parallel$ may be generated by a deficit of charge density with respect to the Goldreich-Julian density -- the so called gaps \citep[e.g.][]{rud75,aro79}. Magnetar magnetospheres are widely believed to be current carrying and twisted \citep[e.g.][]{tho02,bel09}. A high-density pair plasma is expected to fill the magnetosphere so that no global $E_\parallel$ is expected \citep{tho02}, cf. \cite{wad20}. Additional mechanisms are needed to generate $E_\parallel$. One possibility is that a strong $E_\parallel$ can be induced as Alvf\'en waves reach a critical radius where charge starvation occurs \citep{kum20,lu20}. Particle acceleration occurs and an FRB can be generated. The balance between acceleration and radiation cooling requires 
\be
NeE_{\parallel}c\sim \eta N^2\frac{2e^2c\gamma^4}{3\rho^2},
\label{eq:balance}
\ee
where $N$ is the electron number in a coherent bunch, and $\eta \leq 1$ is a coherence factor we introduced. In previous estimations, $\eta=1$ has been assumed so that the radiation power of curvature radiation is $N^2$ times than that of a single electron. This is strictly speaking the case when $N$ electrons are regarded as a point source. Considering realistic bunches in three-dimensional scales, its radiation would be somewhat suppressed due to incoherence, leading to $\eta<1$ (see detailed discussions in \citet{yan18}).

For a pair plasma, the existence of the $E_\parallel$ makes electrons and positrons decouple and separate from each other. This would lead to two-stream instability which facilitates the formation of clumps of particles (see detailed discussions in \citep{kum17,kum20}). In the meantime, a Coulomb field is generated due to pair separation. For an order of magnitude treatment, we approximate the positron and electron clumps as point sources with a separation of $\Delta$, and assumes that the pair clumps keep balance under electric field acceleration and radiation cooling. Without loss of generality, we assume an anti-parallel rotator so that the Goldreich-Julian charge density is positive in the polar region. The Alfv\'en waves will induce an outward $E_\parallel$. Considering this $E_\parallel$ and the secondary Coulomb field due to charge separation, the balance condition of both clumps can be written as
\be
E_\parallel-\frac{N_-e}{\Delta^2}&\sim&\eta N_+\frac{2e\gamma^4}{3\rho^2},~~~~~{\rm for~positron~clump}, \label{positron}\\
\frac{N_+e}{\Delta^2}-E_\parallel&\sim&\eta N_-\frac{2e\gamma^4}{3\rho^2},~~~~~{\rm for~electron~clump}, \label{electron}
\ee
where $N_+$ and $N_-$ are the particle numbers in the positron and electron clumps, respectively. Eliminating $E_\parallel$ in the above equations, one gets
\be
\Delta&\sim&\left(\frac{3}{2\eta\mathcal{M}}\right)^{1/2}\frac{\rho}{\gamma^2}\nonumber\\
&\sim&12~{\rm cm}~\mathcal{M}_4^{-1/2}\eta^{-1/2}\rho_9\gamma_3^{-2}
\ee
where $\mathcal{M}=(N_++N_-)/(N_+-N_-)$ is the pair multiplicity, and $N_+>N_-$ is assumed. In order to keep balance, $E_\parallel\sim N_-e/\Delta^2\sim N_+e/\Delta^2$ is required.
The particle number in a coherent bunch is
\be
N&\sim&\pi \mathcal{M}n_{\rm GJ} \Delta_\perp^2\Delta\nonumber\\
&\sim&6\times10^{19}\mathcal{M}_4 B_{p,14} P_0^{-1}r_9^{-2}\nu_9^{-1}\Delta_1
\ee
where $n_{\rm GJ}=(B_p/Pec)(r/R)^{-3}$ is the Goldreich-Julian density, $\Delta_\perp\sim\sqrt{r\lambda}$ is the maximum transverse size for a bunch emitting coherent radiation. The parallel electric field is required to be
\be
E_\parallel\sim\frac{Ne}{\Delta^2}\sim5\times10^8~{\rm V~cm^{-1}}N_{20}\Delta_{1}^{-2}.
\ee
This value is greater than that estimated by \cite{kum17} for the same parameters, since the existence of screening electric field due to charge separation raises the demand of $E_\parallel$. With these parameters, the isotropic equivalent luminosity is given by
\be
L_{\rm iso}&\sim& \eta N^2\gamma^4\frac{2e^2c\gamma^4}{3\rho^2}\nonumber\\
&\sim&5\times10^{37}~{\rm erg~s^{-1}}\eta N_{20}^2\gamma_3^8\rho_9^{-2}
\ee
where the factor of $\gamma^4$ is attributed to the radiation beaming effect (within a cone of half opening angle $1/\gamma$) and the relativistic propagation effect (by a factor of $\gamma^2$) \citep{kum17}. This is consistent with the isotropic luminosity of FRB 200428.

In the model invoking Alvf\'en-wave-induced $E_\parallel$ \citep{lu20}, the FRB duration is determined by shear wave propagation inside the magnetar crust, i.e., $\tau\sim R/v\sim3~{\rm ms}$ for the wave speed of $v\sim0.01c$.
The typical frequency of Alvf\'en waves may be $\nu_{\rm A}\sim (10^3-10^5)~{\rm Hz}$, and the $E_\parallel$ in the charge starvation region would oscillate with a frequency of $\sim \nu_{\rm A}$ \citep{kum20}. The pair-separation process delineated above would repeat itself within the millisecond duration of the FRB.  One may estimate that there are approximately $\tau \nu_{\rm A}\sim(3-300)$ oscillations to contribute to the observed FRB emission.

At last, we check whether an FRB produced this way can propagate in the pair plasma inside the magnetosphere.
We consider that the pair plasma is streaming relativistically with $\gamma_s$, and has an average spread of the background distribution in the plasma rest frame $K'$, i.e.  $\left<\gamma\right>\sim\gamma/\gamma_s$. In the $K'$ frame, the plasma frequency is 
\be
\omega_p'&=&\sqrt{\frac{4\pi e^2 \mathcal{M}n_{\rm GJ}}{\gamma_s m_e}}\nonumber\\
&\simeq&5\times10^7~{\rm rad~s^{-1}}\mathcal{M}_4^{1/2}B_{p,14}^{1/2}P_0^{-1/2}r_9^{-3/2}\gamma_{s,2}^{-1/2},
\ee
According to the two-stream instability, in the laboratory frame the longitudinal size of a typical
clump is 
\be
l\sim\frac{c}{\gamma_s\omega_p'}\simeq6~{\rm cm}~\mathcal{M}_4^{-1/2}B_{p,14}^{-1/2}P_0^{1/2}r_9^{3/2}\gamma_{s,2}^{-1/2},
\ee
which is also consistent with the above discussion about the pair separation.
On the other hand, in the $K'$ frame the Larmor frequency is
\be
\omega_B'=\frac{eB_p}{m_ec}\left(\frac{r}{R}\right)^{-3}\simeq1.8\times10^{12}~{\rm rad~s^{-1}}B_{p,14}r_9^{-3},
\ee
and the FRB frequency is 
\be
\omega'=2\pi\nu/\gamma_s\simeq6\times10^7~{\rm rad~s^{-1}}\gamma_{s,-2}^{-1}.
\ee 
Thus, $\omega_p'\sim\omega'\ll\omega_B'$ is satisfied. 
Since the polarized direction of curvature radiation is in the trajectory plane, the curvature radiation photons should be O-mode in the emission region. In a magnetized pair plasma, the transparency condition for O-mode photons is \citep{raf19}
\be
\omega'&>&\frac{\omega_p'}{\sqrt{\left<\gamma\right>}}\sin\theta_B\nonumber\\
&\simeq&1.5\times10^6~{\rm rad~s^{-1}}\mathcal{M}_4^{1/2}B_{p,14}^{1/2}P_0^{-1/2}r_9^{-3/2}\gamma_{3}^{-1/2}\theta_{B,-1},\nonumber\\
\ee
where $\theta_B$ is the angle between the field line and photon momentum direction.
Since $\theta\ll1$ for the photons generated by curvature radiation, the pair plasma is transparent for the curvature radiation close to the emission region. As the photons propagating outwards, the plasma frequency $\omega_p'$ would decrease (although $\theta$ increases slightly). The transparent condition is therefore always satisfied along the trajectory of wave propagation.

\section{Coherent radiation from the separated electron/positron bunches}\label{spectrum}

\cite{yan18} calculated the coherent curvature radiation spectra of electron-positron pair bunches and derived a typical $S_\nu\propto\nu^{2/3}$ spectral shape in the low energy regime. Such a spectral shape corresponds to a relatively wide spectrum, which may be in conflict with the non-detection of low-frequency FRBs so far \citep[e.g.,][]{tin15,rav19}. In the following, we improve the calculations by introducing the spatial separation of electron-positron pairs.

We calculate coherent radiation directly from the acceleration of charged particles.
We assume that there are $N$ charged particles moving along a trajectory $\bm{r}(t)$.
The energy radiated per unit frequency interval per unit solid angle is given by \citep[e.g.,][]{jac75}
\be
\frac{dI}{d\omega d\Omega}=\frac{\omega^2}{4\pi^2c}\left|\int_{-\infty}^{+\infty}\sum_j^Nq_j\bm{n}\times(\bm{n}\times\bm{\beta}_j)e^{i\omega(t-\bm{n}\cdot\bm{r}_j(t)/c)}dt\right|^2,\nonumber\\\label{multiemission}
\ee
where $q_j$ is the corresponding charge, $j$ represents the identifier of each charged particle, $\omega$ is the observed angle frequency, $\bm{n}$ is the unit vector between the electron and the observation point, and $\bm{\beta}=\dot{\bm{r}}(t)/c$ is the dimensionless velocity.

\begin{figure*}[]
\centering
\includegraphics[width = 0.4\linewidth , trim = 70 80 100 60, clip]{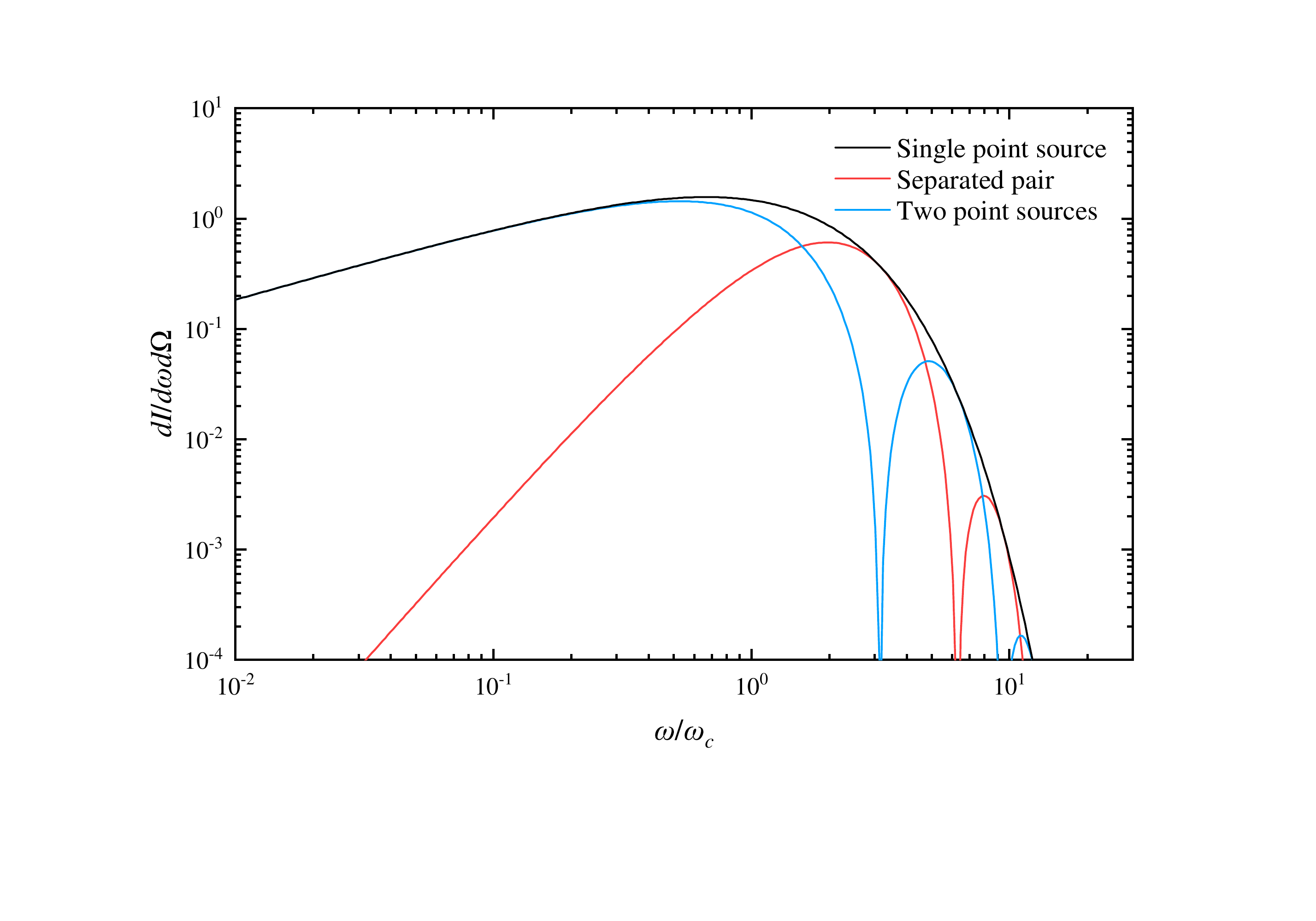}
\includegraphics[width = 0.4\linewidth , trim = 70 80 100 60, clip]{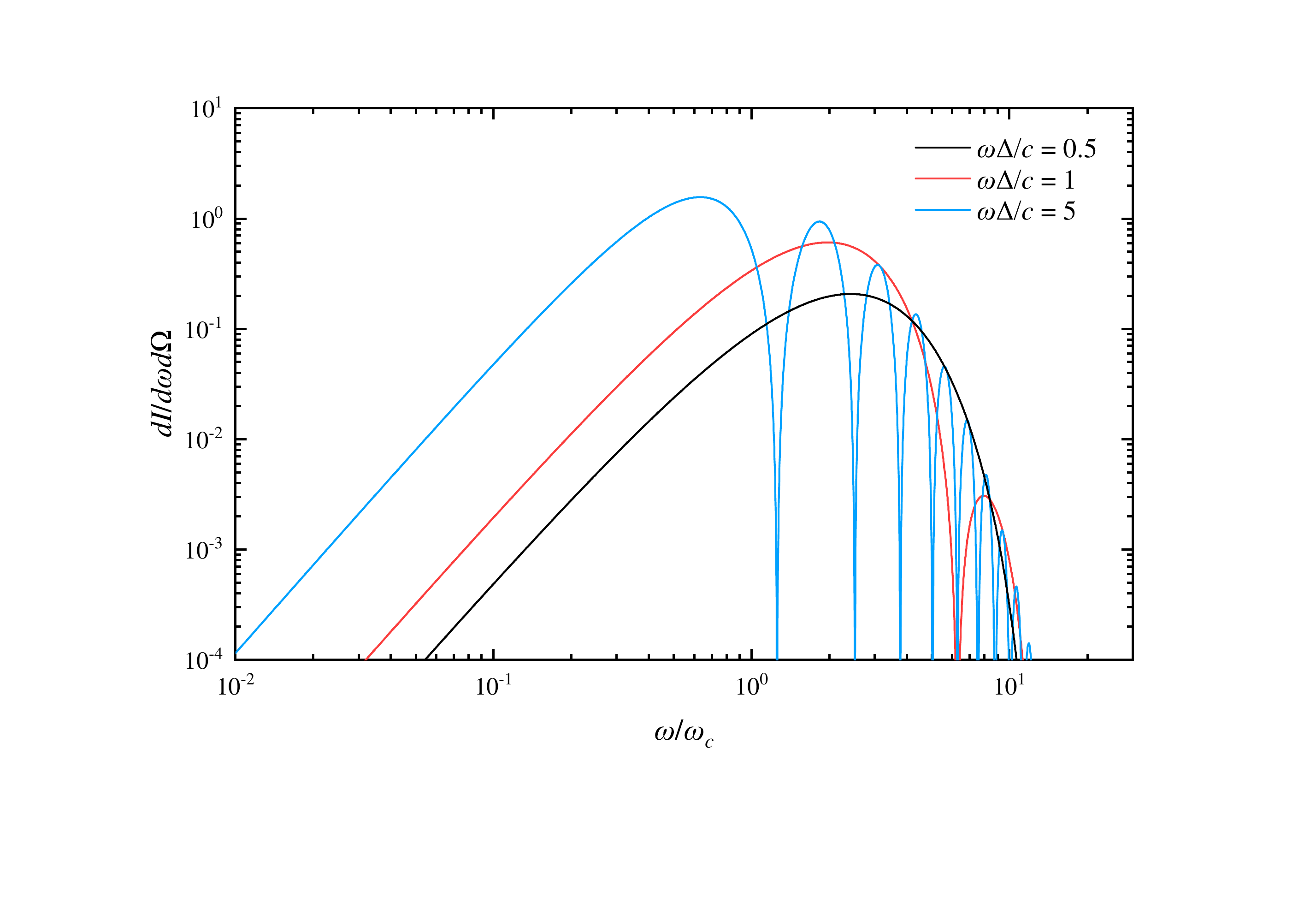}
\caption{Left panel: Coherent curvature radiation spectra for different bunches: a single point source bunch (black line), from two charged-separated clumps with opposite signs (red line), and from two charge-separated clumps with the same sign (blue line). The unit of $dI/d\omega d\Omega$ is arbitrary. Right panel: Coherent curvature radiation spectra for a pair of charge-separated clumps with different separation lengths. The black, red and blue lines correspond to $\omega\Delta/c=0.5,1,5$, respectively.}\label{fig2}
\end{figure*}

\begin{figure*}[]
\centering
\includegraphics[width = 0.4\linewidth , trim = 70 80 100 60, clip]{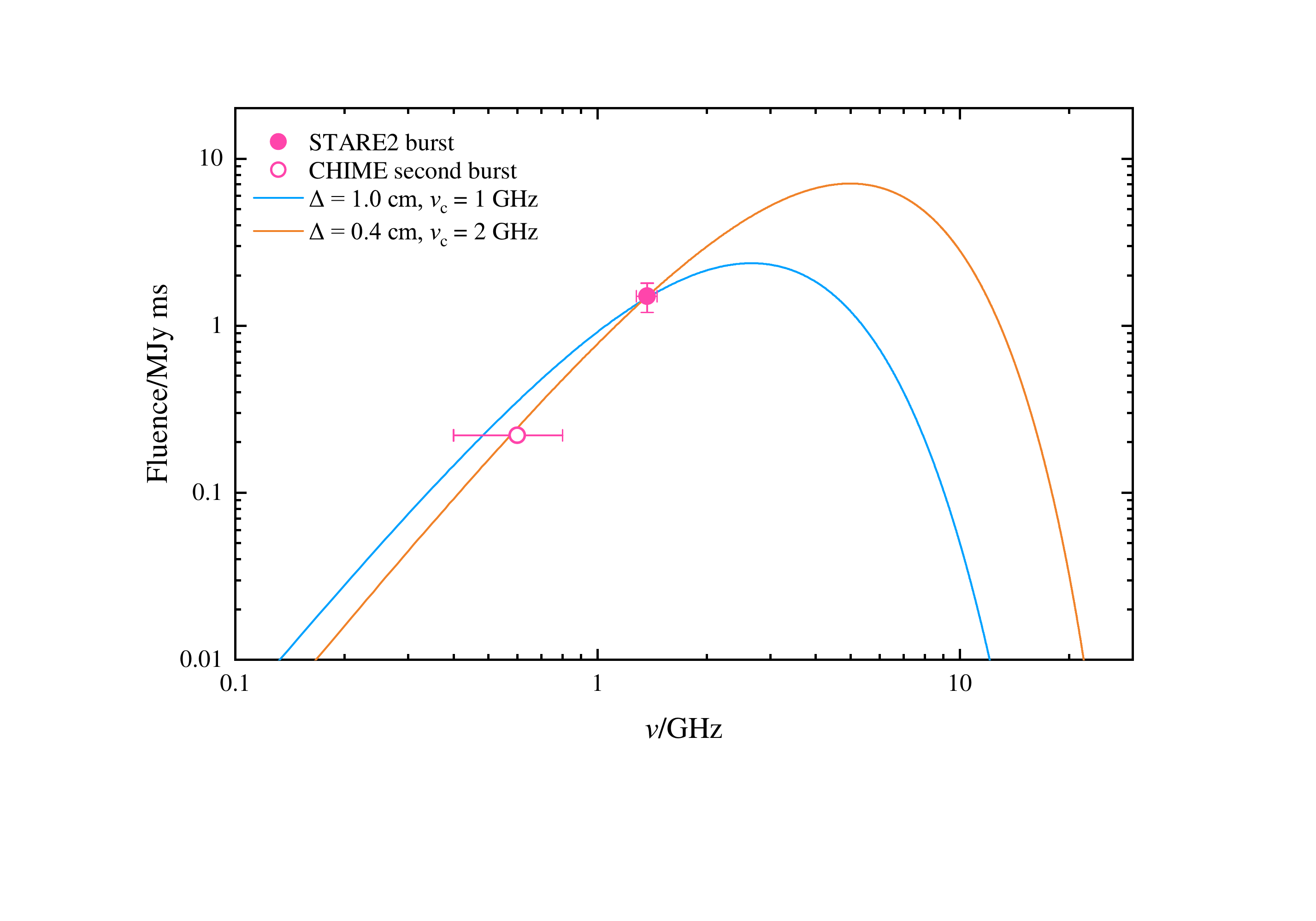}
\includegraphics[width = 0.4\linewidth , trim = 70 80 100 60, clip]{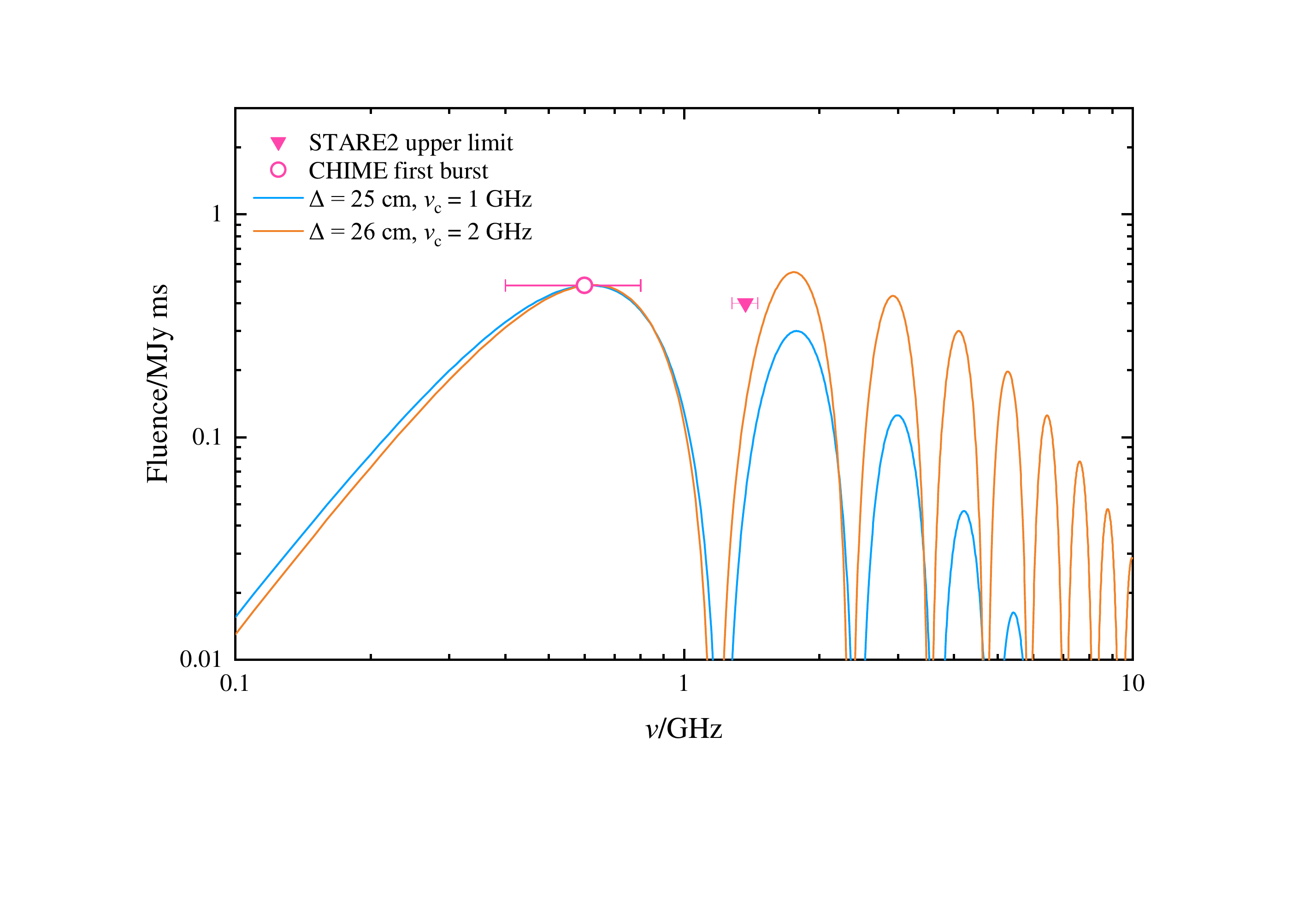}
\caption{The application of this model to FRB 200428. Left panel: The spectrum derived from the CHIME second burst and the STARE2 burst. The blue line corresponds to $\Delta=1.0~{\rm cm}$ and $\nu_c=1~{\rm GHz}$. The orange line corresponds to $\Delta=0.4~{\rm cm}$ and $\nu_c=2~{\rm GHz}$. Right panel: The spectrum derived from the CHIME first burst and the STARE2 upper limit. The blue line corresponds to $\Delta=25~{\rm cm}$ and $\nu_c=1~{\rm GHz}$. The orange line corresponds to $\Delta=26~{\rm cm}$ and $\nu_c=2~{\rm GHz}$.}\label{fig3} 
\end{figure*}

We consider the coherent emission from a pair of charge-separated clumps, as shown in Figure \ref{fig1}. For simplicity, we assume both the electron clump and positron clump as point sources with a separation $\Delta$. We take the electron/positron number in each clump as $N$, then the coherent radiation from the pair of clumps can be calculated by
\be
\frac{dI_{(N)}}{d\omega d\Omega}&=&\frac{N^2e^2\omega^2}{4\pi^2c}\left|\int_{-\infty}^{+\infty}\bm{n}\times(\bm{n}\times\bm{\beta})e^{i\omega(t-\bm{n}\cdot\bm{r}(t)/c)}dt\right|^2\nonumber\\
&\times&\left|1-e^{-i\omega(\bm{n}\cdot\Delta/c)}\right|^2.
\ee
This equation can be written as the radiation of a single electron multiplied by a coherent factor  $\left|1-e^{-i\omega(\bm{n}\cdot\Delta/c)}\right|^2N^2=2\left[1-\cos\left(\omega\bm{n}\cdot\Delta/c\right)\right]N^2$, i.e.
\be
\frac{dI_{(N)}}{d\omega d\Omega}=2\left[1-\cos\left(\frac{\omega\bm{n}\cdot\Delta}{c}\right)\right]N^2\frac{dI_{(1)}}{d\omega d\Omega}, \label{key}
\ee
where the radiation from a single electron satisfies $dI_{(1)}/d\omega d\Omega\propto\omega^{2/3}\exp(-\omega/\omega_c)$ \citep[e.g.,][]{yan18}, and $\omega_c=3\gamma^3c/2\rho$ is the critical frequency. We consider that the observed energy reaches the maximum value when the line of sight is parallel to the trajectory plane, i.e., $\bm{n}\cdot\bm{\Delta}=\Delta$. For $\omega\Delta/c\ll1$, one has
\be
\frac{dI_{(N)}}{d\omega d\Omega}\propto\omega^{8/3}~~~~~{\rm for}~\omega\ll\omega_l\ll\omega_c,
\ee
where 
\be
\omega_l\sim c/\Delta.
\ee 
We can see that the low-frequency spectrum is much harder than $S_\nu\propto\nu^{2/3}$, which appears a narrow spectrum compared with the classical curvature radiation.

On the other hand,  in order to make the radiation from the charge separated clump pair coherent, the condition $\Delta/\rho\ll\theta_c$ needs to be satisfied, where $\theta_c\sim(3c/\omega\rho)^{1/3}$ is the emission angle in the $\omega\ll\omega_c$ regime. Therefore, the upper limit of the coherent frequency is given by
\be
\omega_m\sim\left(\frac{\rho}{\Delta}\right)^2\omega_l.
\ee
Electromagnetic waves with $\omega\gg\omega_m$ would not be coherent between the two clumps, even though they could be coherent within each clump individually. 
It is worth checking whether two adjacent clump pairs are coherent. 
For the Alvf\'en wave with frequency of $\nu_{\rm A}\sim(10^3-10^5)~{\rm Hz}$, the separation between the two pairs is $L\sim c/\nu_{\rm A}$, giving the maximum coherent frequency 
\be
\nu_{M}\sim\frac{c\rho^2}{2\pi L^3}\sim\frac{\rho^2\nu_{\rm A}^3}{2\pi c^2}\simeq0.2~{\rm GHz}~\rho_9^2\nu_{\rm A,4}^3.
\ee 
Therefore, the radiation from two adjacent clump pairs are essentially incoherent.

In the left panel of Figure \ref{fig2}, we plot the coherent curvature radiation spectra for several different bunches; a single point source bunch (black line), from two charged-separated clumps with opposite signs (red line), and from two charge-separated clumps with the same sign (blue line).  
First, we compare the case of one bunch and two bunches with the same sign of charge. Due to the spatial distribution of the charged sources, some narrow spectral structures appear \citep{kat18,yan18}. 
However, in general the complete spectrum for two bunches is still wide, with $S_\nu\propto\nu^{2/3}$ at low frequencies. 
On the other hand, if the two bunches have opposite charges, as expected for charge separation in an external $E_\parallel$, 
the low-frequency radiation is suppressed and the final spectrum becomes narrow.
In the right panel of Figure \ref{fig2}, we plot the coherent curvature radiation spectra of a pair of clumps with opposite charges for different separations. We can see that the spectral structure becomes progressively more complicated as the separation increases. On the other hand, the peak intensity also increases with $\Delta$. This is because as the two clumps are close, the opposite charges tend to cancel out each other to suppress coherence. In any case, the low-frequency spectral index remains 8/3, maintaining a narrow spectrum.

For FRB 200428 from SGR J1935+2154, the CHIME burst shows an average fluence of $0.48~{\rm MJy~ms}$ for the first burst component and $0.22~{\rm MJy~ms}$ for the second burst component in the frequency band $(400-800)~{\rm MHz}$ \citep{CHIME2020}. The STARE2 burst shows an average fluence of $1.5\pm0.3~{\rm MJy~ms}$ at frequency band of $(1281-1468)~{\rm MHz}$.
Assuming that the STARE2 bursts corresponds to the second CHIME burst component, the upper limit on the first CHIME burst component is $0.4~{\rm MJy~ms}$ in the STARE2 frequency band \citep{STARE2020}. 

The spectral feature of FRB 200428 can be interpreted within the framework of our model.  As shown in the left panel of Figure \ref{fig3}, the observational spectrum constructed from the CHIME and STARE2 data for the second burst component is consistent of the low-frequency spectrum predicted by Eq.(\ref{key}), with a spectral index of $8/3$. For the critical frequency of $\nu_c=1~{\rm GHz}$ and $\nu_c=2~{\rm GHz}$, the pair separation is required to be $\Delta=1.0~{\rm cm}$ and $\Delta=0.4~{\rm cm}$, respectively. For given observational data, the larger the critical frequency, the smaller the required separation between the clump pair. 
On the other hand, for the first burst component, the observed fluence decreases as frequency increases (see Figure 1 in \citet{CHIME2020}). Meanwhile, STARE2 did not detect this component and only gives an upper limit.  The constructed spectrum for this component can be accommodated by our model assuming that the CHIME band is around the first peak frequency. As shown in the right panel of Figure \ref{fig3}, 
the data are consistent with the model for different values of the critical frequency. For $\nu_c=1~{\rm GHz}$ and $\nu_c=2~{\rm GHz}$, the clump separation is required to be $\Delta=25~{\rm cm}$ and $\Delta=26~{\rm cm}$, respectively. The smaller the critical frequency, the lower the fluence of high-frequency oscillations.

\section{Summary}\label{discussion}

Prompted by the association of the two bursts of FRB 200428 with the two X-ray peaks in the lightcurve of its X-ray counterpart \citep{HXMT2020}, we further develop the magnetospheric model of FRBs \citep{kum17,yan18} by introducing charge separation of pairs in a parallel electric field due to charge starvation of an Alvf\'en wave \citep{kum20,lu20}. By calculating coherent emission from first principles, we obtain a narrow spectrum with low-frequency spectral index 8/3. This model is found to be able to interpret the observed spectra of the two components of FRB 200428 \citep{CHIME2020,STARE2020}. The model may also give interpretations to other FRBs that have evidence of narrow spectra \citep[e.g.][]{spi16}.

\acknowledgments
We thank Pawan Kumar, Wenbin Lu, and a referee for valuable comments and discussions.
This work is partially supported by the National Natural Science Foundation of China under grant No. 11725314 and the National Basic Research Program of China under grant No. 2014CB845800.

\end{document}